\documentclass[preprint,12pt]{elsarticle}

\usepackage{amssymb}
\usepackage[brazil, english]{babel}
\usepackage[utf8]{inputenc}
\usepackage{graphicx}
\usepackage{epsfig}
\usepackage{amssymb}
\usepackage{amsmath} 

\journal{Journal of Geometry and Physics}

\begin{document}

\begin{frontmatter}


\author{Eduardo Folco Capossoli\corref{cor1}}
\ead{eduardo\_capossoli@cp2.g12.br}

\title{Dynamical Versions to the Holographic Softwall Model: Using Type IIB Superstring Backgrounds via AdS/CFT Correspondence in Hadronic Physics}

 \author[label1,label2]{}
 \address[label1]{Departamento de F\'{\i}sica / Mestrado Profissional
em Pr\'aticas da Educa\c c\~ao B\'asica (MPPEB), Col\'egio Pedro II  , 20.921-903 - Rio de Janeiro-RJ - Brazil}
 \address[label2]{Instituto de F\'{\i}sica, Universidade Federal do Rio de Janeiro, 21.941-972 - Rio de Janeiro-RJ - Brazil}



\begin{abstract}
 In this paper we will show some applications achieved within type IIB superstring backgrounds since the emergence of the AdS/CFT correspondence.
This correspondence relates a super Yang-Mills (SYM) theory, with strong coupling, and a type IIB superstring theory, with weak coupling. The use of the AdS/CFT correspondence allows us to investigate many aspects of hadronic physics described by QCD outside the perturbative regime, for example,
glueball masses and the Regge trajectories related to the pomeron and the odderon. Throughout the text we will deal with the AdS/CFT correspondence within the two dynamical versions to the holographic softwall model.

\end{abstract}


\end{frontmatter}


\section{Introduction}
\label{}

The Anti-de Sitter/Conformal Field Theory (AdS/CFT) correspondence or duality, also known by string/gauge or gauge/gravity duality or yet holography was proposed by Juan Maldacena in 1988 \cite{mal} and brought various perspectives for the hadronic physics outside the perturbative regime in the sense that one can relate an abstract type IIB superstring theory to a SYM theory. 

Correspondences or dualities are not new in the history of Physics. These correspondences relate two physical theories, generally distinct from one another, through certain characteristics belonging to both theories. Just for example, we recall two very well-known dualities in Physics. The first one is the duality between the quantum Sine-Gordon model and the massive Thirring model \cite{ref10, ref11} and the second one is the electric-magnetic duality or Seiberg's duality \cite{Seiberg:1994pq}.

The major difference between AdS/CFT correspondence and those mentioned before is that the first two relate two quantum field theories to each other while the AdS/CFT correspondence relates a quantum field theory in a $d$-dimensional space and a theory of supergravity in a curved $D$-dimensional space, with $D> d$.

More generally, however, with high level of mathematical abstraction, it can be said that the AdS/CFT correspondence relates a superstring theory or M-theory on certain background which can be described by $AdS_d \times {\cal M}^{D-d}$, where $AdS_d$ is $d$-dimensional anti-de Sitter space and ${\cal M}^{D-d}$ is some compactification of the $D-d$-dimensional space, with a conformal field theory (CFT) on the AdS boundary. One can note, if $D=10$ one has a superstring theory. In case of $D=11$, one has a M-theory.

In a more concrete view, AdS/CFT correspondence can be seen as a correspondence or duality between a conformal SYM, since there is no dynamic in the $\beta$ function, with extended supersymmetry $({\cal N} = 4)$, symmetry group given by $SU(N)$ with $N \rightarrow \infty$, in a flat $(3+1)$ dimensional Minkowski spacetime and a type IIB superstring theory in a curved $10$-dimensional spacetime, which can be mathematically described as the $AdS_5 \times S^5$, which in a low energy limit, can be associated to a theory of supergravity. Furthermore, there are many well known references dealing with AdS/CFT as one can see in \cite{Gubser:1998bc,Witten:1998qj,Witten:1998zw,Aharony:1999ti}, for instance.

The use of the AdS/CFT correspondence allows us to investigate many aspects of the hadronic physics describe by the QCD outside the perturbative regime. QCD is the best known theory to describe the strong interactions. Amongst other characteristics of QCD, there is one called confinement, meaning that in the infrared limit, i.e, for low energies or large distances, quarks and gluons are bound to each other strongly, the strong coupling $g \gg 1$,  inaccessible to the perturbative approach. Such calculations involve bound states as the glueball masses, and consequently its related Regge trajectories, are  features of the non-perturbative regime. 

Glueball states are bound states of gluons predicted by QCD, but not detected so far and characterised by $J^{PC}$, where $J$ is the total angular momentum, and $P$ and $C$ are the $P-$parity (spatial inversion)  and  the $C-$parity (charge conjugation) eigenvalues, respectively.

Regge trajectories are well known approximate linear relations between total angular momenta $(J)$ and the square of the masses $(m)$, such as:
\begin{equation}
J(m^2) \approx  \alpha' \, m^2 + \alpha_0 \, ,
\end{equation}
\noindent with $\alpha_0$ and $\alpha'$ constants.

In order to make the comparison between our results and the results coming from other approaches easier, we will provide the most known Regge trajectory for the soft pomeron \cite{Donnachie:1984xq, Donnachie:1985iz}, given by:
\begin{equation}\label{land}
J(m^2) \approx 0.25 \, m^2  + 1.08\,,
\end{equation}
\noindent where the masses throughout this work are expressed in GeV. The pomeron is related to the even spin glueball, with $P=C=+1$. In a $J \times m^2$ plane known as Chew-Frautschi plane, the masses of glueball states lie on the pomeron Regge trajectory.

Of course, there are also other models to describe the pomeron, providing Regge trajectories pretty close to Eq.\eqref{land}, as one can see for instance in \cite{Cudell:2001ii, Levin:1998pk}.

On the other hand, for odderon, now related to the odd spin glueball with $P=C=-1$, there are also many models to describe it, such as, isotropic lattice \cite{Meyer:2004jc}, anisotropic lattice \cite{Chen:2005mg}, relativistic many body model  \cite{LlanesEstrada:2005jf} and the non-relativistic constituent model \cite{LlanesEstrada:2005jf} for which: 
\begin{equation}\label{nr_odderon}
J(m^2) \approx  0.18 \, m^2  + 0.25 \,,
\end{equation}
etc. We are going to compare our results obtained for the Regge trajectory of the odderon, with this trajectory.

As mentioned before the AdS/CFT implies a super conformal field theory, and thus, cannot be used directly to tackle QCD, since QCD is not a conformal theory. So, one must break the conformal invariance, and after this, one can construct phenomenological models that describe (large $N$) QCD approximately. These models are known as AdS/QCD models.

Some proposals appeared in order to deal with the conformal invariance, such as ``Witten black hole" \cite{Witten:1998zw} and the introduction of an IR hard cutoff a certain value $z_{max}$ of the holographic coordinate $z$ and just considering a slice of $AdS_5$ space in the region $0 \leq z  \leq z_{max}$, with some appropriate boundary conditions \cite{Polchinski:2001tt, Polchinski:2002jw, BoschiFilho:2002vd,  BoschiFilho:2002ta}. In these two last works emerges the idea of the hardwall model, as this model is known nowadays. There are many results in order to study even and odd glueball state masses, as well as Regge trajectories associated to the pomeron and the odderon within the hardwall model, as one can see in the following references \cite{BoschiFilho:2005yh, Capossoli:2013kb, Rodrigues:2016cdb}.

For our purposes, in this work  we will focus on another approach  known as softwall model, in its dynamic version, to break the conformal invariance and investigate the hadronic physics, as can be seen in the following Sections. 

The softwall model arises from the need to break the conformal invariance and one has to introduce in the action of the fields a decreasing exponential factor of the dilatonic field that represents a soft IR cutoff. The original softwall model was proposed  in \cite{Karch:2006pv} to study vector mesons, and subsequently extended to  glueballs \cite{Colangelo:2007pt}, to other mesons and baryons \cite{Forkel:2007tz} and even to study the deep inelastic scattering \cite{Capossoli:2015sfa}. The main feature of this model is to produce linear Regge trajectories. 

As discussed in \cite{BoschiFilho:2012xr, Li:2013oda, Capossoli:2015ywa} 
the Regge trajectories for glueballs coming from the original softwall model although linear are not in agreement with lattice data. In particular in reference \cite{FolcoCapossoli:2016ejd} suggested an $AdS_5$ mass renormalisation in order to get a unified treatment for both scalar and high even spin glueballs. Due to this, modifications were proposed in the original softwall model, as can be seen in the following Sections.

This work is organized as follows: In Section \ref{dsw} we will present a dynamical modification in the holographic softwall model in order to calculate the masses of the even and odd spin glueballs as well as the Regge trajectories related to pomeron and odderon. Also in this Section, we will explore the analytically solvable version of this model. In the Section \ref{ano} we will present a modification in the dynamical softwall model, now taking into account the anomalous dimension related to the QCD beta function and numerically solve, also in order to get the masses of the even and odd spin glueballs and the Regge trajectories related to pomeron and odderon. Finally in Section \ref{con} we will present our conclusions and last comments.

\section{Dynamical Modification in the Holographic Softwall Model} \label{dsw}

As mentioned in the previous Section, the references \cite{BoschiFilho:2012xr, Li:2013oda, Capossoli:2015ywa} showed that the softwall model does not seem to be working well for glueball states because both masses and Regge trajectories are not in agreement with those found in the literature. In this Section, we will present a modification to the softwall model.

The dynamical modification in the holographic softwall model is based on the dilaton field becoming dynamical satisfying the Einstein's equations, and the metric structure is also consistently solved by Einstein's equations, and both cases are in five dimensions.

To do this, let us start writing the $5D$ action for the graviton-dilaton action in the string frame:
\begin{equation}\label{acao_corda}
S = \frac{G_5^{-1}}{16 \pi } \int d^5 x \sqrt{-g_s} \; e^{-2\Phi(z)} ({\cal R}_s + 4 \partial_M \Phi \partial^M \Phi - V^s_G(\Phi))
\end{equation}
\noindent where $G_5$ is Newton's constant in five dimensions, $g_s$ is the determinant of the metric tensor in the $5-$dimensional space, $\Phi(z) = k z^2$ is the dilatonic field, where $k \sim \Lambda^2_{QCD}$ and $V^s_G(\Phi)$ is the dilatonic potential. All of these parameters are in the string frame, so the metric tensor has the following form:
\begin{equation}\label{g_s}
ds^2 = g^s_{MN} dx^M dx^N = b^2_s(z)(dz^2 + \eta_{\mu \nu}dx^\mu dx^\nu); \; \; \;b_s(z) \equiv e^{A_s(z)}.
\end{equation}
\noindent with $M,N = 0,1,2,3,4; \; \mu, \nu = 0,1,2,3,$ and  $\eta_{\mu \nu} =$ diag $(-1, 1, 1, 1)$ the metric of the four-dimensional Minkowski space. 
After a Weyl rescaling, from the string frame to the Einstein frame, one can write Eq.(\ref{acao_corda}) as:
\begin{equation}\label{acao_einstein}
S = \frac{1}{16 \pi G_5} \int d^5 x \sqrt{-g_E} \; (R_E -\frac{4}{3} \partial_M \Phi \partial^M \Phi - V^E_G(\Phi))\;, 
\end{equation}
\noindent where
\begin{equation}\label{weyl}
 g^E_{MN} = g^s_{MN}e^{-\frac{2}{3}\Phi}\;; \qquad V^E_G = e^{\frac{4}{3}\Phi}V^s_G\;.
\end{equation}

The equations of motion from \eqref{acao_einstein}, can be writte as:
\begin{equation}\label{eq_mov_e_2_1}
 -A''_E + A'^2_E - \frac{4}{9}\Phi'^2  = 0\;;
\end{equation}
\begin{equation}\label{eq_mov_e_2_2}
 \Phi'' + 3A'_E \Phi' - \frac{3}{8}e^{2A_E}\partial_\Phi V^E(\Phi) = 0\;,
\end{equation}
\noindent where we defined $\Phi'=\partial \Phi/ \partial z$, $A'=\partial A/ \partial z$ and
\begin{equation}\label{redef}
b_E (z) = b_s(z)e^{-\frac{2}{3}\Phi(z)} = e^{A_E(z)}\;; \qquad A_E(z) = A_s(z) - \frac{2}{3}\Phi(z)\;.
\end{equation}
\noindent Solving Eqs. (\ref{eq_mov_e_2_1}) and (\ref{eq_mov_e_2_2}) for the quadratic dilaton background, $\Phi(z)=kz^2$, one finds:
\begin{equation}\label{sol_eq_mov_e_2_1}
 A_E(z) = \log{\left( \frac{R}{z} \right)} - \log{\left(_0F_1\left(\frac 54, \frac{\Phi^2}{9}\right)\right)}\;, 
\end{equation}
\noindent and
\begin{equation}\label{sol_eq_mov_e_2_2}
 V^E_G(\Phi) = -\frac{12 ~ _0F_1(\frac14, \frac{\Phi^2}{9})^2}{R^2} + \frac{16 ~ _0F_1(\frac 54, \frac{\Phi^2}{9})^2\, \Phi^2}{3 R^2}\;,
\end{equation}
where $_0F_1(a,z)$ is the Kummer confluent hypergeometric function. 
Using (\ref{redef}) and (\ref{sol_eq_mov_e_2_1}), one can note that the warp factor in the string frame is
\begin{equation}\label{redef_2}
 A_s(z) = \log{\left( \frac{R}{z} \right)}  + \frac{2}{3}\Phi(z) - \log{\left[_0F_1\left(\frac 54, \frac{\Phi^2}{9}\right)\right]}\,, 
\end{equation}
\noindent which means that the metric (\ref{g_s}) is a deformed AdS space. Using \eqref{weyl} one has

\begin{equation}\label{vs}
 V^s_G(\Phi) =\exp\{-\frac 43 \Phi\} \left[ -\frac{12 ~ _0F_1(1/4, \frac{\Phi^2}{9})^2}{R^2} + \frac{16 ~ _0F_1(5/4, \frac{\Phi^2}{9})^2 \Phi^2}{3 R^2}\right]
\end{equation}

\noindent so that this potential generates the desired quadratic dilaton where $R$ is the AdS radius.

Returning to string frame, the $5D$ action for the scalar glueball field ${\cal G}$ is given by  \cite{Colangelo:2007pt}:
\begin{equation}\label{acao_ori_soft}
S = \int d^5 x \sqrt{-g_s} \; \frac{1}{2} e^{-\Phi(z)} [\partial_M {\cal G}\partial^M {\cal G} + M^2_{5} {\cal G}^2]
\end{equation}
\noindent which leads to the following equation of motion:
\begin{equation}\label{eom_1}
\partial_M[\sqrt{-g_s} \;  e^{-\Phi(z)} g^{MN} \partial_N {\cal G}] - \sqrt{-g_s} e^{-\Phi(z)} M^2_{5} {\cal G} = 0\,.
\end{equation}
Representing the scalar field through a $4d$ Fourier transform ${\cal \tilde{G}}(q,z)$ and performing a change of function ${\cal \tilde{G}} = \psi (z) e^{\frac{B(z)}{2}}$, where $B(z) = \Phi(z) - 3A_s(z) $, one gets the following $1d$ Schr\"odinger-like equation
\begin{equation}\label{equ_5}
- \psi''(z) + \left[ \frac{B'^2(z)}{4}  - \frac{B''(z)}{2} + M^2_{5} \left( \frac{R}{z}\right)^2  e^{4kz^2/3} {\cal A}^{-2} \right] \psi(z) = - q^2 \psi(z)
\end{equation}
\noindent or explicitly for the quadratic dilaton $\Phi(z)= k z^2$: 
\begin{equation}\label{equ_7}
- \psi''(z) + \left[ k^2 z^2 + \frac{15}{4z^2}  - 2k + M^2_{5} \left( \frac{R}{z}\right)^2  e^{4kz^2/3}\right] \psi(z) = (- q^2 )\psi(z),
\end{equation}
with ${\cal A}$ = $_0F_1(5/4, \frac{\Phi^2}{9})$. This equation was solved numerically in \cite{Li:2013oda,Capossoli:2016kcr, Capossoli:2016ydo}.

Since, for our purposes, in this Section, we are interested in analytical solutions, then we will use the action \eqref{acao_ori_soft} with metric tensor (\ref{g_s}) and $\Phi(z)$ still given by $k z^2$ but with the function $A_s(z)$ replaced by: 
\begin{equation}\label{am}
{{A}}_M(z) = \log{\left( \frac{R}{z} \right)}  + \frac{2}{3}\Phi(z),
\end{equation}

One can conclude by looking at (\ref{g_s}) and (\ref{am}) that this modification produces a deformation in the original $AdS_5$, meaning that this dynamical softwall model is no longer $AdS_5$. Of course, now we are dealing with an asymptotically $AdS_5$ space, as can be seen for the UV limit or $z\rightarrow 0$, one has $A_M(z)|_{(z\rightarrow 0 )}\propto \log \left( \frac{R}{z} \right)$ .

The Schr\"odinger-like equation \eqref{equ_7} has an effective potential given by:
$${\cal V}(z) = \left[ k^2 z^2 + \frac{15}{4z^2} - 2k + M^2_{5} \left( \frac{R}{z}\right)^2  e^{4kz^2/3} \right]\,.$$
This is still not exactly solvable so we expand the exponential in the last term in the brackets and just retain terms up to first order in the parameter $k$. In fact, we could retain terms up to second order in $k$ without breaking exact solvability, but this contribution would not modify significantly our subsequent analysis. This procedure gives us the equation 
\begin{equation}\label{equ_7_1_new}
- \psi''(z) + \left[ k^2 z^2 + \frac{15}{4z^2}  - 2k + M^2_{5} \left( \frac{R}{z}\right)^2   + \frac{4 kz^2}{3}M^2_{5} \left( \frac{R}{z}\right)^2\right] \psi(z) = (- q^2 )\psi(z).
\end{equation}
which is exactly solvable and represents the dynamical and analytical softwall model that we consider here. From the eigenenergies and by associating $-q^ 2_n$ with the square of the masses of the 4D glueball states, one has:
\begin{equation}\label{adsw_1}
m_n^2 = \left[ 4n + 2\sqrt{4 +M^2_{5} R^2} + \frac{4}{3}R^2M^2_{5} \right]k; \;\;\;\; (n=0, 1, 2, \cdots). 
\end{equation}

For the lightest scalar glueball $0^{++}$ is dual to the fields with zero mass $(M^2_5 = 0 )$ in the $AdS_5$ space, Eq.(\ref{adsw_1}) becomes:
\begin{equation}\label{adsw}
m_n^2 = \left[ 4n +4 \right] k \,.
\end{equation}
The results obtained from \eqref{adsw_1} for the masses of the lightest scalar glueball $(n = 0)$ and its radial excitations $(n=1, 2, \cdots)$, using $k = 0.85$ GeV$^2$ are presented in Table  \ref{t1}. 
\begin{table}[h]

\centering
\begin{tabular}{|c|c|c|c|c|c|c|c}
\hline
 &  \multicolumn{4}{c|}{Glueball States $J^{PC}$}  & \\  
\cline{2-5}
 & $0^{++}$ & $0^{++*} $ & $0^{++**}$ & $0^{++***}$ & $k$ \\ \hline
 $n$ & 0 & 1 & 2 & 3 & \\ 
\hline \hline
\, $m_n$ \,                                   
&\, 1.84\, &\, 2.61 \,&\, 3.19 \,& \, 3.69 \, & \, 0.85 \, \\ \hline
\end{tabular}
\caption{\em Masses expressed in GeV for the glueball states $J^{PC}$ of the the lightest scalar glueball and its radial excitations from dynamical and analytical softwall model using Eq.(\ref{adsw}) for $k = 0.85$ GeV$^2$.}
\label{t1}
\end{table}

The values found for the masses of the lightest scalar glueball and its radial excitations are in agreement with those found in the literature from lattice calculations, as one can see in \cite{Meyer:2004jc, Chen:2005mg, Morningstar:1999rf, Lucini:2001ej}.

In order to deal with higher spin glueballs one can recall the AdS/CFT correspondence dictionary who tells us how to relate the operator in the gauge theory with fields in the $AdS_{5} \times S^5$ space. The conformal dimension $\Delta$ of a boundary operator is given by:
\begin{equation}\label{dim_delta}
\Delta = 2 + \sqrt{4 + R^2 M^2_{5}}
\end{equation}
For a pure SYM theory defined on the boundary, one has that the scalar glueball state $0^{++}$ is represented by the operator ${\cal O}_4$, given by:
\begin{equation}\label{fmn}
{\cal O}_4 = Tr(F^{\mu\nu}F_{\mu \nu})
\end{equation}
which has conformal dimension $\Delta = 4$. So, the lightest scalar glueball $0^{++}$ is dual to the fields with zero mass $(M^2_{5} = 0 )$ in the $AdS_5$ space, as mentioned before.

After this explanation, we will apply an approach following \cite{deTeramond:2005su} where the glueball operator with spin $\ell$ could be obtained by the insertion of symmetrised covariant derivatives in the operator ${\cal O}_{4} = F^2$, such that:
\begin{equation}
 {\cal O}_{4+ \ell} = FD_{\{\mu_1 \cdots} {D_{\mu_\ell\}}}F
 \end{equation}
 \noindent with conformal dimension $\Delta = 4 + \ell$.

This approach was used within holographic hardwall model in two cases. The first one, to calculate the masses of even glueball states $0^{++}$, $2^{++}$, $4^{++}$, $6^{++}, \cdots$ and to obtain the corresponding pomeron Regge trajectory \cite{BoschiFilho:2005yh, Rodrigues:2016cdb}. The second one, was used to calculate the masses of even glueball states $1^{--}$, $3^{--}$, $5^{--}$, $7^{--}, \cdots$ and to obtain the corresponding odderon Regge trajectory \cite{Capossoli:2013kb}.

For even spin glueball states after the insertion of symmetrised covariant derivatives, and using \eqref{dim_delta}, one has:
\begin{equation}
M^2_{5}R^2 = \ell(\ell+4)\,; \qquad ({\rm even}\, \ell)\,.
\end{equation}
\noindent Plugging this result in Eq.(\ref{adsw_1}), one gets:
\begin{equation}\label{adsw_1_even}
m_n^2 = \left[  4n + 2\sqrt{4 +\ell(\ell+4)} + \frac{4}{3}\ell(\ell+4) \right]k\,; \qquad ({\rm even}\, \ell).
\end{equation}
and for the particular cases of non-excited states $(n=0)$, one has:
\begin{equation}\label{kkssparn}
m_0^2 = \left[  2\sqrt{4 +\ell(\ell+4)} + \frac{4}{3}\ell(\ell+4) \right]k\,; \qquad ({\rm even}\, \ell).
\end{equation}

In the case of odd spin glueballs, following, the operator ${\cal O}_6$ that describes the glueball state $1^{--}$ is given by:
\begin{equation}
 {\cal O}_{6} =SymTr\left( {\tilde{F}_{\mu \nu}}F^2\right),
 \end{equation} 
\noindent and inserting the symmetrised covariant derivatives one has:
\begin{equation}
{\cal O}_{6 + J} = SymTr\left( {\tilde{F}_{\mu \nu}}F D_{\lbrace\mu_1 \cdots} D_{\mu_\ell \rbrace}F\right),
\end{equation}

\noindent with conformal dimension $\Delta = 6 + J$ and spin $1+\ell$. Then, for the case of the odd spin glueball states, using again \eqref{dim_delta}, one finds:
\begin{equation}
M^2_{5}R^2 = (J+6)(J+2)\,; \qquad ({\rm odd}\, \ell), 
\end{equation}

\noindent Plugging this result in Eq.(\ref{adsw_1}), one gets:
\begin{equation}\label{adsw_1_odd}
m_n^2 = \left[  4n + 2\sqrt{4 +(J+6)(J+2)} + \frac{4}{3}(J+6)(J+2) \right]k\,; \qquad ({\rm odd}\, \ell).
\end{equation}

\noindent One can read for the non-excited odd spin glueball states $(n=0)$
\begin{equation}\label{kkssimpn}
m_0^2 = \left[ 2\sqrt{4 +(J+6)(J+2)} + \frac{4}{3}(J+6)(J+2) \right]k\,; \qquad ({\rm odd}\, \ell).
\end{equation}


In Table \ref{t5}, we will present the values for masses for the even spin glueball states, from \eqref{kkssparn} and in the Table \ref{t6},  the values for the masses for the odd spin glueball states, from \eqref{kkssimpn}. For both calculation was used $k= 0.2$ GeV$^2$.
\begin{table}[h]

\centering
\begin{tabular}{|c|c|c|c|c|c|c|c|}
\hline
 &  \multicolumn{6}{c|}{Glueball States $J^{PC}$}  & \\  
\cline{2-7}
 & $0^{++}$ & $2^{++} $ & $4^{++}$ & $6^{++}$ & $8^{++}$ & $10^{++}$  & $ k $ \\
\hline \hline
Masses                                   
&\, 0.89\, &\, 2.19 \,&\, 3.30 \,& \, 4.38 \, &\, 5.44 &\, 6.49 \, & \, 0.20 \, \\ \hline 
\end{tabular} 
\caption{\em Masses expressed in GeV for the glueball states $J^{PC}$ with even $J$ from the dynamical and analytical softwall model using Eq.(\ref{kkssparn}) with $k= 0.2$ GeV$^2$.}
\label{t5}
\end{table}

\begin{table}[h]

\centering
\begin{tabular}{|c|c|c|c|c|c|c|c|}
\hline
 &  \multicolumn{6}{c|}{Glueball States $J^{PC}$}  & \\  
\cline{2-7}
 & $1^{--}$ & $3^{--} $ & $5^{--}$ & $7^{--}$ & $9^{--}$ & $11^{--}$  & $ k $ \\
\hline \hline
Masses                                   

&\, 2.82\, &\, 3.94 \,&\, 5.03 \,& \, 6.11 \, &\, 7.19&\, 8.26\, & \,  0.20 \, \\ \hline
\end{tabular}
\caption{\em Masses expressed in GeV for the glueball states $J^{PC}$ with odd $J$ from SW using Eq.(\ref{kkssimpn}) and $k= 1$ and 2 GeV$^2$ and from the modified SW using Eq.(\ref{adsw_1_odd}) and $k= 0.2$ GeV$^2$.} 
\label{t6}
\end{table}

From these results presented in Table \ref{t5}, one can derive the Regge trajectory for even spin glueball states, which one can be associated to the pomeron, such as:
\begin{equation}\label{rtpadsw}
J(m^2) = (0.23 \pm 0.02) \, m^2 + (0.8 \pm 0.5)
\end{equation}
\noindent The errors for the slope and the intercept come from the linear fit. This Regge trajectory is in agreement with the one presented in \eqref{land}.

In the same way, using the results from Table \ref{t6}, one can derive the Regge trajectory for odd spin glueball states, which one can be associated to the odderon, such as:
\begin{equation}\label{rtoadsw}
J(m^2) = (0.17 \pm 0.01) \, m^2 + (0.4 \pm 0.4)\;.
\end{equation}
\noindent The errors for the slope and intercept come from the linear fit. This Regge trajectory for the odderon is in agreement with the one presented in \eqref{nr_odderon}, within the nonrelativistic constituent model.

\section{Dynamical Corrections to the Anomalous Holographic Softwall Model} \label{ano}

In this Section we will calculate numerically the masses of  higher, even and odd, spin glueball states, and construct the Regge trajectories related to pomeron and odderon but now taking into account the anomalous dimensions from  a chosen QCD beta function,  namely, beta function with an IR fixed point at finite coupling, in addition to dynamical corrections in the holographic softwall model in same way that was done in the previous Section.

The references  \cite{Gursoy:2007cb} introduced the idea of using QCD beta functions to get an interesting UV behaviour for the softwall model modified by convenient superpotentials for the dilaton field. In particular in \cite{BoschiFilho:2012xr}, the authors took into account the anomalous dimensions, also related to QCD beta functions, and obtained the masses only for the scalar glueball and its radial excitations in agreement with those presented in the literature. 

As we are dealing with dynamical corrections in the softwall model, we will follow the same steps used in Section \ref{dsw}. Then let us recall the metric tensor that will be used:
\begin{equation}\label{redef_2_1}
 A_s(z) = \log{\left( \frac{R}{z} \right)}  + \frac{2}{3}\Phi(z) - \log{\left[_0F_1\left(\frac 54, \frac{\Phi^2}{9}\right)\right]}\,, 
\end{equation}

\noindent and the action for the scalar glueball field ${\cal G}$ that also will be used: 
\begin{equation}\label{acao_ori_soft_1}
S = \int d^5 x \sqrt{-g_s} \; \frac{1}{2} e^{-\Phi(z)} [\partial_M {\cal G}\partial^M {\cal G} + M^2_{5} {\cal G}^2].
\end{equation}
The dilatonic field still remains as $\Phi = k z^2$.

After some calculation as done in the Section \ref{dsw} we obtained this Schr\"odinger-like equation:
\begin{equation}\label{equ_7_1}
- \psi''(z) + \left[ k^2 z^2 + \frac{15}{4z^2}  - 2k + M^2_{5} \left( \frac{R}{z}\right)^2  e^{4kz^2/3}\right] \psi(z) = (- q^2 )\psi(z).
\end{equation}

Also recalling the AdS/CFT dictionary, the classical (non-anomalous) conformal dimension $\Delta_{\rm{class.}}$ of a super Yang-Mills (SYM) scalar operator is given by:
\begin{equation}\label{di}
\Delta_{\rm{class.}} = 2 + \sqrt{4 + R^2 M^2_{5}}\;. 
\end{equation}
Therefore, one can write:
\begin{equation}\label{dim}
R^2 M^2_{5} = \Delta_{\rm{class.}}( \Delta_{\rm{class.}} - 4) \;. 
\end{equation}

The SYM is a conformal theory, so the beta function vanishes and the conformal dimensions has no anomalous contributions, therefore they keep only their classical dimension. 

On the other side, within the QCD approach, the scalar glueball operator has full dimension given from the trace anomaly of the energy-momentum tensor \cite{Narison:1988ts,Gubser:2008yx}, so that:
\begin{equation}\label{beta1}
T^{\mu}_{\mu} = \frac{\beta(\alpha)}{16 \pi \alpha^ 2} Tr F^2 + {\rm fermionic \;\;terms}
\end{equation}

\noindent and the beta function can defined as:
\begin{equation}\label{beta2}
\beta(\alpha(\mu) )\equiv \frac{d \alpha(\mu)}{d \ln(\mu)},
\end{equation}
where $\mu$ is a renormalisation scale, $\alpha \equiv g_{YM}^2 /4 \pi$ and $g_{YM}$ is the Yang-Mills coupling constant. 
The fermionic part in (\ref{beta1}) can be disregarded because only the operator $Tr F^2$ is relevant for our purposes. 
Besides, the scaling behaviour for a generic operator can be written as:
\begin{equation}\label{beta3}
\Delta_{\cal O} = - \frac{d {\cal O}}{d \ln \mu}.
\end{equation}
It is appropriate to mention that the full dimension $\Delta_{\cal O}$ also can be represented by sum of the classical dimension $\Delta_{\rm{class.}}$ and the anomalous dimension $\gamma(\mu)$, then one has:
\begin{equation}\label{beta4}
\Delta_{\cal O} = \Delta_{\rm{class.}} + \gamma(\mu).
\end{equation}

Particularly, for the case of the scalar glueball operator, inserting  Eq.(\ref{beta1}), disregarding  fermionic part, in (\ref{beta3}), we obtain:
\begin{equation}
\Delta_{T^{\mu}_{\mu}} \left( \frac{\beta(\alpha)}{8 \pi \alpha^ 2} Tr F^2 \right)  = - (\beta'(\alpha) - \frac{2}{\alpha} \beta(\alpha) - \Delta_{F^2}) \frac{\beta(\alpha)}{8 \pi \alpha^ 2} Tr F^2\;,
\end{equation}

\noindent where the prime represents the derivative with respect to $\alpha$.

Then, the scalar glueball operator $Tr F^2$ has the full dimension:
\begin{equation}\label{beta6}
\Delta_{F^2} = 4 + \beta'(\alpha) - \frac{2}{\alpha} \beta(\alpha)
\end{equation}

Using the 't Hooft coupling $\lambda \equiv N_C g_{YM}^2 = 4 \pi N_C \alpha$, one gets
\begin{equation}\label{beta7}
\Delta_{F^2} = 4 + \beta'(\lambda) - \frac{2}{\lambda} \beta(\lambda)
\end{equation}

\noindent where the prime represents the derivative with respect to $\lambda$ and the beta function is given by:
\begin{equation}
\beta(\lambda(\mu)) \equiv \frac{d \lambda(\mu)}{d \ln(\mu)}.
\end{equation}

As our concern is about higher spin glueballs, for even and odd spins, and to get their corresponding Regge trajectories related to the pomeron and to the odderon, we will use the same procedure done in Section \ref{dsw}, that means we will insert symmetrised covariant derivatives in a given operator with spin $\ell$ in order to raise the total angular momentum. In the particular case of the operator ${\cal O}_4 = F^2$, one gets once again:
\begin{equation}\label{4+J}
{\cal O}_{4 + J} = FD_{\lbrace\mu_1 \cdots} D_{\mu_\ell \rbrace}F,
\end{equation}
\noindent with conformal dimension $\Delta_{\rm{class.}} = 4 + \ell$ and spin $\ell$.  

In a similar way,  one can write the full dimension $\Delta^{even\,J}_{T^{\mu}_{\mu}} = 4 + J$, 
and now Eq.(\ref{beta7}) can be written as:
\begin{equation}\label{beta8}
\Delta^{even\, J}_{F^2} = 4 + \ell + \beta'(\lambda) - \frac{2}{\lambda} \beta(\lambda).
\end{equation}
Using (\ref{dim}), the full dimension for a glueball state with higher even spin $\ell$, taking into account the beta function is:
\begin{equation}\label{dfullp}
R^2M^2_{5} = \Delta^{even\, \ell}_{F^2}  (\Delta^{even\, \ell}_{F^2}  -4)
\end{equation}
or explicitly:
\begin{equation}\label{r2par}
R^2M^2_{5} = \left[  4 + \ell + \beta'(\lambda) - \frac{2}{\lambda} \beta(\lambda)\right] \left[ \ell + \beta'(\lambda) - \frac{2}{\lambda} \beta(\lambda)\right]\,; \qquad ({\rm even}\, \ell)\,.
\end{equation}

One has to replace \eqref{r2par} in the Schr\"odinger-like equation \eqref{equ_7_1} to get the masses for even glueball states.

On the other hand, for odd spin glueballs, as also shown in Section \ref{dsw} the operator ${\cal O}_6$ that describes the glueball state $1^{--}$ is given by:
\begin{equation}
 {\cal O}_{6} =SymTr\left( {\tilde{F}_{\mu \nu}}F^2\right),
 \end{equation} 

\noindent and after the insertion of symmetrised covariant derivatives one gets:
\begin{equation}\label{6+J}
{\cal O}_{6 + J} = SymTr\left( {\tilde{F}_{\mu \nu}}F D_{\lbrace\mu_1 \cdots} D_{\mu_\ell \rbrace}F\right),
\end{equation}

\noindent with conformal dimension $\Delta_{\rm{class.}} = 6 + \ell$ and spin $1+\ell$. 

Then one can write the full dimension $\Delta^{odd\, \ell}_{T^{\mu}_{\mu}} = 6 + \ell$\,, 
and now Eq.(\ref{beta7}) becomes:
\begin{equation}\label{beta8_1}
\Delta^{odd\, \ell}_{F^2} = 6 + \ell + \beta'(\lambda) - \frac{2}{\lambda} \beta(\lambda).
\end{equation}
Using (\ref{dim}), one can write the full dimension for a glueball state with higher odd spin $\ell$, taking into account the beta function:
\begin{equation}\label{dfulli}
R^2M^2_{5} = \Delta^{odd\, \ell}_{F^2}  (\Delta^{odd\, \ell}_{F^2}  -4)
\end{equation}
and explicitly:
\begin{equation}\label{r2impar}
R^2M^2_{5} = \left[  6 + \ell + \beta'(\lambda) - \frac{2}{\lambda} \beta(\lambda)\right] \left[ 2 + \ell + \beta'(\lambda) - \frac{2}{\lambda} \beta(\lambda)\right]\,; \qquad ({\rm odd}\, \ell).
\end{equation}
One has to replace \eqref{r2impar} in \eqref{equ_7_1} to get the masses of the odd spin glueball states.

At this moment, let us discuss about  the QCD beta function chosen for this work whose was proposed  in \cite{Alanen:2009na}:
\begin{equation}\label{beta11}
\beta(\lambda) = - b_0 \lambda^2 \left[ 1 - \frac{\lambda}{\lambda_{\ast}}\right] \;;\;\;\; {\rm for}\;\;\; \lambda_{\ast} > 0\;.
\end{equation}

This beta function fulfil  necessary IR and UV requirements, meaning that for the IR fixed point $\lambda = \lambda_{\ast}$ this beta function vanishes. Moreover, it reproduces the perturbative $\beta(\lambda) \sim - b_0 \lambda^2$ at $1-$ loop order in the ultraviolet and behaves as $\beta (\lambda) \sim + \lambda^3$ at large coupling. 

Since one can relate the holographic or radial coordinate $z$ of the $AdS_5$ space with $\mu^{-1}$ where $\mu$ was defined as the renormalisation group scale, one can write the relationship between the beta function and coordinate $z$, given by:
\begin{equation}\label{beta10}
\beta(\lambda(\mu)) = \mu \frac{d \lambda(\mu)}{d \mu} \Rightarrow  \beta(\lambda(z)) = - z \frac{d \lambda(z)}{dz},
\end{equation}

\noindent where the integration constant will be fixed by $\lambda(z) \equiv \lambda_0$ at a particular energy scale $z_0$.

Eq. (\ref{beta10})  can also be solved exactly for this beta function, so that:
\begin{equation}\label{beta12}
\lambda(z) = \frac{\lambda_{\ast}}{1 + W\left(\left( \frac{z_0}{z}\right)^{b_0 \lambda_{\ast}} \left( \frac{\lambda_{\ast} - \lambda_0}{\lambda_0}\right)  \exp^{\frac{\lambda_{\ast} - \lambda_0}{\lambda_0}}\right) }
\end{equation}

\noindent where $W(z)$ is again the Lambert function and $\lambda(z_0) = \lambda_0$ fixes the integration constant. 
This equation leads to the expected QCD asymptotic behaviour at short distances when $z$ is close to the boundary $(z\to 0)$:
\begin{equation}
\lambda(z) \sim - 1/(b_0 \ln z).
\end{equation}

Finally, replacing (\ref{beta11}) and (\ref{beta12}) in Eqs. (\ref{r2par}) and (\ref{r2impar}), solving numerically the Schr\"odinger-like equation \eqref{equ_7_1}  and using suitable values for $k$, $\lambda_0$ and $\lambda_{\ast}$, one can get the masses of even and odd glueball states, respectively. 

The results obtained for the masses of even and odd glueball states are presented in Table \ref{t7} and Table \ref{t8}, respectively.

\begin{table}[!h]
\centering
\begin{tabular}{|c|c|c||c|c|c|c|c|c|} \hline  
   \multicolumn{3}{|c||}{Parameters} 
&   \multicolumn{6}{c|}{Glueball States $J^{PC}$ Masses}  \\  \hline  
\cline{4-9}   
 \hline    $k$ & $\lambda_0$ & $\lambda_{\ast}$  & $0^{++}$ & $2^{++} $ & $4^{++}$ & $6^{++}$ & $8^{++}$ & $10^{++}$   \\
\hline \hline                        
 \hline                             
  $ 0.09$ & $10.5$ & $350$  & 0.79 & 2.13 & 3.28 &  4.39  & 5.48  & 6.57   \\ \hline                             
 \end{tabular} 
\caption{\em  Masses {\rm (GeV)}  for the glueball states $J^{PC}$ with even $\ell$ with $P=C=+1$ calculated numerically from dynamical corrections to the anomalous holographic softwall model using \eqref{equ_7_1} with mass relationship \eqref{r2par} and the beta function with an IR fixed point at finite coupling, \eqref{beta11}, using suitable values for the  parameters  $k$ {\rm (GeV$^{2}$)},  $\lambda_0$ and $\lambda_{\ast}$ (dimensionless).}
\label{t7}
\end{table}

\begin{table}[h]
\centering
\begin{tabular}{|c|c|c||c|c|c|c|c|} \hline  
   \multicolumn{3}{|c||}{Parameters} 
&   \multicolumn{5}{c|}{Glueball States $J^{PC}$ Masses}  \\  \hline  
\cline{4-8}   
 \hline    $k$ & $\lambda_0$ & $\lambda_{\ast}$  & $1^{--}$ & $3^{--} $ & $5^{--}$ & $7^{--}$ & $9^{--}$    \\
\hline \hline                        
 \hline                             
  $ 0.09$ & $10.5$ & $350$  & 2.72 & 3.84 & 4.94 & 6.03 & 7.11    \\ \hline                             
 \end{tabular} 
\caption{\em  Masses {\rm (GeV)}  for the glueball states $J^{PC}$ with even $\ell$ with $P=C=-1$ calculated numerically from dynamical corrections to the anomalous holographic softwall model using \eqref{equ_7_1} with mass relationship \eqref{r2impar} and the beta function with an IR fixed point at finite coupling, \eqref{beta11}, using suitable values for the  parameters  $k$ {\rm (GeV$^{2}$)},  $\lambda_0$ and $\lambda_{\ast}$ (dimensionless).}
\label{t8}
\end{table}

From Table \ref{t7} one can derive the following Regge trajectory related to the pomeron:
\begin{equation}\label{pad}
J(m^2) =  (0.23 \pm 0.02) \, m^2 + (0.9  \pm 0.5)   \,,
\end{equation}
in agreement with the one presented in \eqref{land}. The errors for the slope and the intercept come from the linear fit.

In the same way, from Table \ref{t8} one can derive the following Regge trajectory related to odderon:
\begin{equation}\label{oad}
J(m^2) = (0.18 \pm 0.01)\, m^2 + (0.1 \pm 0.4) \, ,
\end{equation}
in agreement with the one presented in \eqref{nr_odderon}. The errors for the slope and the intercept come from the linear fit.

\section{Conclusions} \label{con}

In this work we presented two examples how the type IIB superstring theory via AdS/CFT correspondence can be used to investigate the hadronic physics away from the perturbative regime.

The approach used is this work was based on the dynamical versions of the holographic softwall model. The first one, analytically solvable and the second one, numerically solvable and taking into account the corrections from anomalous dimension. Both approaches give results for the masses of higher even and odd spin glueball, as well as the Regge trajectories to associated to the pomeron and the odderon compatible with those found in the literature.

\section{Acknowledgements}
E.F.C. is partially supported by PROPGPEC-CPII. E.F.C. would like to thank Henrique Boschi-Filho for his suggestions and comments.


\end{document}